\documentclass[fleqn,usenatbib]{mnras}

\usepackage{newtxtext,newtxmath}

\usepackage[T1]{fontenc}

\DeclareRobustCommand{\VAN}[3]{#2}
\let\VANthebibliography\thebibliography
\def\thebibliography{\DeclareRobustCommand{\VAN}[3]{##3}\VANthebibliography}

\usepackage{graphicx}	
\usepackage{amsmath}	
\usepackage{mathrsfs}
\usepackage{hyperref}
\usepackage{pdflscape}
\usepackage{float}
\usepackage{here}
\usepackage{longtable}
\usepackage[export]{adjustbox}
\usepackage{array}

\newcommand{\xmm}{\textit{XMM-Newton }}

\newcommand{\NXS}{$\sigma_{\rm NXS}^2$\,}
\newcommand{\mass}{$M_{\rm BH}$\,}
\newcommand{\lx}{$L_{\rm 2-10}$\,}
\newcommand{\eddr}{$\lambda_{\rm Edd}$\,}

\title[BASS-XL]{BASS-XL: X-ray variability properties of unobscured Active Galactic Nuclei}

\author[A. Tortosa et al.]{Alessia Tortosa $^{1,2}$\thanks{E-mail: alessia.tortosa@mail.udp.cl, alessia.tortosa@inaf.it},
Claudio Ricci $^{1,3,4}$, Patricia Arévalo$^{5}$, Michael J. Koss$^{6}$, Franz E. Bauer$^{7,8,9,10}$,
\newauthor
Benny Trakhtenbrot$^{11}$, Richard Mushotzky$^{12,13}$,
Matthew J. Temple$^{1}$, Federica Ricci$^{14,2}$,
\newauthor
Alejandra Rojas Lilayu $^{1,15}$, Taiki Kawamuro$^{16}$, Turgay Caglar $^{17}$, Tingting Liu$^{18}$, Fiona Harrison$^{19}$, 
\newauthor
Kyuseok Oh$^{20,21}$, Meredith Clark Powell$^{22}$, Daniel Stern$^{23}$, Claudia Megan Urry$^{24}$.\\
$^{1}$Instituto de Estudios Astrof\'isicos, Facultad de Ingenier\'ia y Ciencias, Universidad Diego Portales, Av. Ej\'ercito Libertador 441, Santiago, Chile\\
$^{2}$ INAF Osservatorio Astronomico di Roma, Via Frascati 33, 00078 Monte Porzio Catone (RM), Italy.\\
$^{3}$Kavli Institute for Astronomy and Astrophysics, Peking University, Beijing 100871, China\\
$^{4}$George Mason University, Department of Physics \& Astronomy, MS 3F3, 4400 University Drive, Fairfax, VA 22030, USA\\
$^{5}$ Instituto de F\iısica y Astronom\iıa, Facultad de Ciencias, Universidad de Valpara\'iso, Gran Bretana N 1111, Playa Ancha, Valpara\'iso, Chile.\\
$^{6}$ Eureka Scientific, 2452 Delmer Street Suite 100, Oakland, CA 94602-3017, USA.\\ 
$^{7}$ Instituto de Astrof\'isica, Facultad de Física, Pontificia Universidad Cat\'olica de Chile, Campus San Joaqu\'in, Av. Vicu\~na Mackenna 4860, Santiago, Chile, 7820436.\\
$^{8}$ Centro de Astroingenier\'ia, Facultad de Física, Pontificia Universidad Cat\'olica de Chile, Campus San Joaqu\'in, Av. Vicuña Mackenna 4860, \\ \,\,\,\,Santiago, Chile, 7820436.\\
$^{9}$ Millennium Institute of Astrophysics, Nuncio Monseñor S\'otero Sanz 100, Of 104, Providencia, Santiago, Chile.\\
$^{10}$ Space Science Institute, 4750 Walnut Street, Suite 205, Boulder, Colorado 80301, USA.\\
$^{11}$ School of Physics and Astronomy, Tel Aviv University, Tel Aviv 69978, Israel.\\
$^{12}$ Department of Astronomy, University of Maryland, College Park, MD 20742, USA.\\
$^{13}$ Joint Space-Science Institute, University of Maryland, College Park, MD 20742, USA.\\
$^{14}$ Dipartimento di Matematica e Fisica, Università degli Studi Roma Tre, via della Vasca Navale 84, 00146 Roma, Italy.\\
$^{15}$ Centro de Astronom\'ia (CITEVA), Universidad de Antofagasta, Avenida Angamos 601, Antofagasta, Chile.\\ 
$^{16}$ RIKEN Cluster for Pioneering Research, 2-1 Hirosawa, Wako, Saitama 351-0198, Japan.\\
$^{17}$ Department of Physics, Southern Methodist University, 3215 Daniel Ave., Dallas, TX 75205, USA.\\
$^{18}$ Department of Physics and Astronomy, West Virginia University, P.O. Box 6315, Morgantown, WV 26506, USA.\\
$^{19}$ Cahill Center for Astronomy and Astrophysics, California Institute of Technology, Pasadena, CA 91125, USA.\\
$^{20}$ Korea Astronomy \& Space Science institute, 776, Daedeokdae-ro, Yuseong-gu, Daejeon 34055, Republic of Korea.\\
$^{21}$ Department of Astronomy, Kyoto University, Kitashirakawa-Oiwake-cho, Sakyo-ku, Kyoto 606-8502, Japan.\\
$^{22}$ Kavli Institute for Particle Astrophysics and Cosmology, Stanford University, 452 Lomita Mall, Stanford, CA 94305, USA.\\
$^{23}$ Jet Propulsion Laboratory, California Institute of Technology, 4800 Oak Grove Drive, MS 169-224, Pasadena, CA 91109, USA.\\
$^{24}$ Yale Center for Astronomy \& Astrophysics and Department of Physics, Yale University, P.O. Box 208120, New Haven, CT 06520-8120, USA.\\
}
\date{Accepted XXX. Received YYY; in original form ZZZ}

\pubyear{2023}

\begin{document}
\label{firstpage}
\pagerange{\pageref{firstpage}--\pageref{lastpage}}
\maketitle

\begin{abstract}
We investigate the X-ray variability properties of Seyfert\,1 Galaxies belonging to the BAT AGN Spectroscopic Survey (BASS). The sample includes 151 unobscured (N$_{\rm H}<10^{22}$\,cm$^{-2}$) AGNs observed with \xmm for a total exposure time of $\sim27$\,Ms, representing the deepest variability study done so far with high signal-to-noise \xmm observations, almost doubling the number of observations analysed in previous works. We constrain the relation between the normalised excess variance and the $2-10$\,keV AGN luminosities, black hole masses and Eddington ratios. We find a highly significant correlation between \NXS and \mass, with a scatter of $\sim 0.85$\,dex. For sources with high \lx this correlation has a lower normalization, confirming that more luminous (higher mass) AGNs show less variability. We explored the \NXS vs \mass relation for the sub-sample of sources with \mass estimated via the "reverberation mapping" technique, finding a tighter anti-correlation, with a scatter of $\sim 0.65$\,dex. We examine how the \NXS changes with energy by studying the relation between the variability in the hard ($3-10$\,keV) and the soft ($0.2-1$\,keV)/medium ($1-3$\,keV) energy bands, finding that the spectral components dominating the hard energy band are  more variable than the  spectral components dominating in softer energy bands, on timescales shorter than 10\,ks.\\
\end{abstract}


\section{Introduction}
\label{sect:intro}
Supermassive Black Holes (SMBHs, $M_{\rm BH}>10^6M_{\odot}$) are ubiquitously found at the center of massive galaxies. Mass accretion onto SMBHs is the mechanism that powers Active Galactic Nuclei (AGNs, \citealt{Salpeter1964}) which are very powerful sources of X-ray radiation, emitting through the entire electromagnetic spectrum. 
Variability is a distinctive feature shared by all classes of AGN, occurring over a wide range of timescales and amplitudes across all the wavelengths (e.g., \citealp{1997ARA&A..35..445U,2004MNRAS.348..783M}). These flux variations can also be accompanied by prominent spectroscopic changes (e.g., \citealp{Ricci:2022}). In the X-ray band, variability is observed on both short \citep[e.g., $<10^3$\,s;][]{2005MNRAS.363..586U,2004MNRAS.348..783M} and long timescales \citep[e.g., years; ][]{2001ASPC..224..205M,2009A&A...504...61I,10.1093/mnrasl/sly025} giving insight into the innermost regions of the AGN. Thus, its study can help us to understand the emission properties of AGNs (e.g., \citealp{1993ARA&A..31..717M,1997ARA&A..35..445U,2014A&ARv..22...72U,2021iSci...24j2557C,2022arXiv220913467D}) and better characterize the growing population of extremely variable AGNs identified in the optical (e.g., \citealp{Lawrence:2016bi,Rumbaugh:2018iv,Trakhtenbrot:2019qz,Shen:2021ws,2022ApJ...939L..16Z,2023MNRAS.518.2938T}) and X-rays (e.g., \citealp{Timlin:2020hz,Ricci:2020,Ricci:2021,Masterson:2022oe}).\\
One method used to study the temporal structure of the variations is the power spectral density (PSD) analysis. If the temporal frequency is $\nu=1/t$, where $t$ is the time, the observed power spectrum is generally modeled as a power-law of the form: $P_{\nu} \propto \nu^{\alpha}$. For short timescales (high frequencies) $\alpha \sim -2$, while for long timescales (low frequencies) $\alpha \sim -1$ \citep{1995MNRAS.273..923P}. The PSD break timescales, $T_B$, can be obtained by fitting a broken power laws to the observed PSD. This parameter has been found to be positively correlated with the black hole mass (\mass; e.g., \citealp{2001MNRAS.324..653L,2002A&A...395..465B,2002MNRAS.332..231U,2003ApJ...598..935M,2004MNRAS.348..207P}). However, Narrow Line Seyfert\,1 (NLS1) galaxies, which typically accrete at very high Eddington ratios ($L_{\rm bol}$/$L_{\rm Edd}=\lambda_{\rm Edd}$; \citealt{2004MNRAS.348..783M}), display a different behaviour, with their break timescales being shorter for a given \mass. To explain this, \citet{2005MNRAS.363..586U} suggested that the break timescales could depend also on a second parameter, such as the accretion rate or the black hole spin.\\
Accurately determining the AGNs power spectra can be difficult, since it requires high-quality data, long exposures and sometimes monitoring campaigns, to extend time coverage that adequately covers relevant PSD frequency ranges that include potential breaks. Given such difficulties, it is common practice to quantify the X-ray variability of AGNs in terms of the so-called normalised excess variance (\NXS, \citealt{1997ApJ...476...70N}). Although it does not contain the same amount of information as the PSD, the normalised excess variance can be used to confirm the PSD results in large samples of AGN, and it also allows the discovery of new correlations between the X-ray variability amplitude and other AGNs physical parameters. The normalised excess variance of AGNs has been widely studied in the past decades, finding that \NXS has a strong dependence on \mass. Using the data from the {\it Advanced Satellite for Cosmology and Astrophysics} ({\it ASCA}), \citet{2001MNRAS.324..653L} and \citet{2003MNRAS.343..164B} found an anti-correlation between the excess variance (on a timescales of $\sim$1\,day) and \mass. \citet{2004MNRAS.348..207P}, using {\it Rossi X-Ray Timing Explorer} ({\it RXTE}) data on much longer timescales ($\sim$300\,days), also found an anti-correlation between these two parameters. \citet{Ponti2012} investigated this relation using high quality {\it XMM-Newton} data on timescales of 10\,ks. They found that the \NXS - \mass relation flattens for masses below $\sim10^6 M_{\odot}$, as confirmed later also by \citet{2015MNRAS.447.2112L} studying a sample of low mass AGNs observed by \xmm. \citet{2022A&A...666A.127A}, using light curves of local Seyfert from the Nuclear Spectroscopic Telescope Array hard X-ray mission (\textit{NuSTAR}), extended the \NXS vs \mass relation to energy band higher than 10\,keV, finding that it is possible to accurately measure the \mass in AGN using the above-mentioned correlation in the $3-10$ and the $10-20$\,keV bands. However, the minimum necessary S/N is $\sim 3$ and duration of the light curves should be $\sim 80 -100$\,ks.\\
Several works suggested that the excess variance is
related to other source properties, such as the X-ray luminosity, \lx, \citep{1986Natur.320..421B, 1997ApJ...476...70N,1999ApJ...524..667T}.
However, studying a sample of 46 AGNs observed by {\it ASCA}, \citet{2004MNRAS.348..207P} found that once the dependence of \NXS from \mass is removed, the correlation between \NXS and \lx is no longer present, implying that the correlation with \lx was associated to the $\sigma_{\rm NXS}^2 \, -\, M_{\rm BH}$ relation. The same effect was recovered by \citet{2005MNRAS.358.1405O}.\\
Past studies of hard X-ray selected AGNs with \textit{Swift}/BAT data, focusing on longterm light curves show that in most of these AGNs a significant variability on months to years timescales is present. In general this variation is not related to changes of the absorption column density but to variations of the power-law continuum \citep{2014A&A...563A..57S}. Moreover, unlike previous studies, no correlation between hard X-ray variability and different properties of the AGNs including luminosity and black hole mass was found \citep{Shimizu_2013}. Also \citet{2023MNRAS.518.4372P}, studying the hard X-ray variability properties of \textit{Swift}/BAT AGNs, show that type 1 AGNs in the 14--150\,keV band, are found to be less prone to harboring deterministic variability than type 2 AGNs on timescales of $\sim15$\,years.\\ 
In this paper, we present the results from an excess variance analysis of a sample of 151 hard X-ray selected, unobscured (N$_H <10^{22}$\,cm$^{-2}$) AGNs using $\sim$500 high signal-to-noise \xmm observations, almost double of the number of observations analysed in previous works (e.g., \citealt{Ponti2012}), for a total of $\sim$27\,Ms exposure time.\\
The paper is organized as follows. Section \ref{sect:sample_reduction} presents the selected sample and the data reduction of the sources of our sample. Section \ref{sect:analysis} describes the timing analysis of the data and the extrapolation of the \NXS together with the analysis of the correlation between \NXS and several physical parameters of the sources. We summarize and conclude the results of our analysis in Section \ref{sect:conclusion}. Standard cosmological parameters (H=70\,km\,s$^{-1} \rm Mpc^{-1}$, $\Omega_{\Lambda}$=0.73 and $\Omega_m$=0.27) are adopted throughout the paper.
\section{The sample and data reduction}
\label{sect:sample_reduction}
\begin{table}
\caption{Summary of the \xmm observations of the sources (OBSID) of our sample together with the {\it Swift} identification name ({\it Swift} ID) and the {\it Swift} identification number (ID). This table is available in its entirety in a machine-readable form in the online journal. A part is shown as guidance for the reader regarding its content.}
\label{tab:obsid}
\begin{tabular}{lll}
\hline
\hline
ID&\textit{Swift}ID&OBSID\\
\hline
\hline
6     & SWIFTJ0006.2+2012    & 0101040701\\
6     & SWIFTJ0006.2+2012    & 0510010701\\
...&\\
16    & SWIFTJ0029.2+1319    & 0783270201\\
34    & SWIFTJ0051.6+2928    & 0903040301\\
36    & SWIFTJ0051.9+1724    & 0801890301\\
39    & SWIFTJ0054.9+2524    & 0301450401\\
39    & SWIFTJ0054.9+2524    & 0841480101\\
43    & SWIFTJ0059.4+3150    & 0312190101\\
61    & SWIFTJ0113.8-1450    & 0147920101\\
73    & SWIFTJ0123.9-5846    & 0101040201\\
73    & SWIFTJ0123.9-5846    & 0721110201\\
...&\\
77    & SWIFTJ0127.5+1910    & 0112600601\\
77    & SWIFTJ0127.5+1910    & 0830551001\\
...&\\
106   & SWIFTJ0206.2-0019    & 0201090201\\
106   & SWIFTJ0206.2-0019    & 0554920301\\
...&\\
\hline
\hline
\end{tabular}
\end{table}
\subsection{The BASS Sample}
\label{sect:sample}
Since its launch in 2004, the Burst Alert Telescope (BAT; \citealp{Barthelmy:2005uq}) on board the {\it Neil Gehrels Swift observatory} \citep{Gehrels:2004dq} has been carrying out an all-sky survey in the $14-195$\,keV band. Our sample consists of all the unobscured ($N_H< 10^{22} \rm cm^{-2}$), radio quiet, type\,1 AGNs belonging to the Swift/BAT AGN Spectroscopic Survey (BASS\footnote{\url{www.bass-survey.com}}) which have public \xmm observations by December 2022.\\
Being unbiased by obscuration up to Comptont-thick levels (N$_{\rm H}>10^{24}$\,cm$^{-2}$, \citealp{Ricci:2015}) and not affected by dust obscuration or star formation, BASS provides an important census of AGNs. It gives a full picture of the bright AGNs in the local Universe, providing the largest available spectroscopic sample of {\it Swift}/BAT ultra-hard X-ray ($14-195$\,keV) detected AGNs \citep{2018ApJS..235....4O}, complementary with {\it Swift}, {\it Chandra}, and \xmm for X-ray broad-band ($0.5-200$\,keV) spectral measurements \citep{Ricci2017}. It also includes extensive multi-wavelength follow-up data, from optical emission \citep{2022ApJS..261....4O}, high spatial resolution near-IR \citep{2017MNRAS.467..540L,2018Natur.563..214K}, mid- and far-IR emission from {\it WISE}, {\it IRAS}, {\it Spitzer}, {\it Akari}, and {\it Herschel} \citep{2017yCat..18350074I,2017MNRAS.466.3161S,Ichikawa_2019} and mm/radio emission \citep{Kawamuro:2022ks,2021ApJS..252...29K,2023ApJ...952L..28R}, giving insight on the sample over the broadest possible spectral range.\\
The first BASS data release (DR1, \citealt{Koss2017}) reported \mass and X-ray properties for all the 838 AGNs from the {\it Swift}/BAT 70-month catalogue \citep{Ricci2017}, while the second BASS data release (DR2, \citealt{2022ApJS..261....1K}) reports more secure and uniformly assessed \mass for 780 unbeamed AGNs from the 70-month catalogue. The masses are estimated from broad Balmer lines and/or "reverberation mapping" technique (RM) for type 1s and masers, and from dynamics and/or velocity dispersions for type 2s \citep{Koss_2022,2022ApJS..261....5M,2022ApJS..261....8R}. Moreover, in the DR2, \eddr ($L_{\rm bol}$/$L_{\rm Edd}$) are computed using the bolometric luminosities calculated from the intrinsic 14--150\,keV luminosities as shown in \citealt{Ricci2017} with a bolometric correction of 8 \citep{2022ApJS..261....1K}. In this work, we considered \lx, \eddr, and \mass values from BASS DR2 \citep{2022ApJS..261....2K}.
\begin{figure*}
	\includegraphics[width=\textwidth]{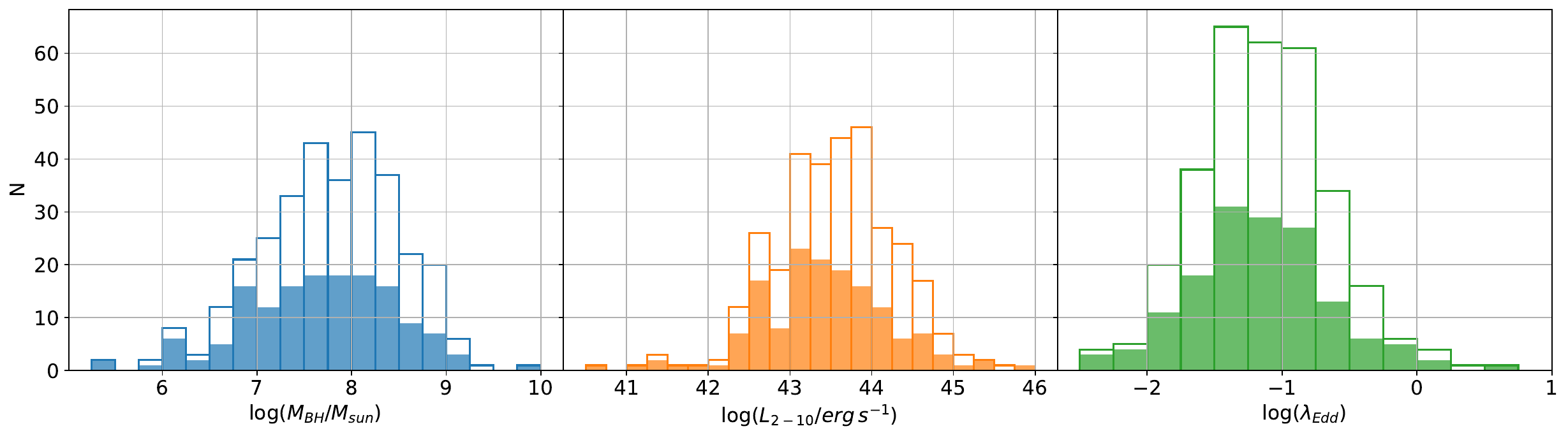}
\caption{Distributions of black hole masses (\mass; left panel), 2--10\,keV luminosities (\lx; middle panel), Eddington ratios (\eddr; right panel) of the sources analysed in this work, compared with the parent BASS sample of unobscured AGNs (empty bars).}
\label{fig:distributions}
\end{figure*}
\subsection{Data Reduction}
\label{sect:data_reduction}
The X-ray Multi-Mirror Mission (\xmm, \citealt{Jansen2001}) high statistics, low background and uninterrupted light-curves obtained for the sources in this sample are crucial to compute the \NXS in large samples of AGN.\\
The sample of BASS unobscured AGNs is composed by 365 sources, with a median redshift of $z_{\rm med}=0.035$ (lower than the parent sample median redshift). Of these, 153 had public \xmm observations as of December 2022. We downloaded all the observations from the \xmm Science Archive, and extracted the EPIC-pn \citep{Struder2001} light curves using the Science Analysis System (\textsc{SAS}) software package (v.18.0.0) \citep{Gabriel2004} and the calibration database \textsc{CALDB 20221102}. The MOS detectors \citep{Turner2001} and the Reflection Grating Spectrometer (RGS, \citealt{denHerder2001}) were not considered because their lower statistics would not significantly improve the quality of the lightcurves.\begin{table*}
\caption{List of the \textit{Swift} Identification number (ID), the \textit{Swift} name (\textit{Swift}ID), the counterpart name, the redshift (z), the hydrogen column density (N$_H$ [cm$^{\rm-2}$]), the mass (\mass in solar masses), the 2--10\,keV luminosity (\lx[$\rm erg\,s^{-1}$]), the bolometric lumunosity ($L_{\rm bol}$[$\rm erg\,s^{-1}$]), the Eddington ratio (\eddr), the normalised excess variance for the light curves binned with 100s (\NXS$_{\rm 100s}$) and 1000s (\NXS$_{\rm 1000s}$)of time binning. This table is available in its entirety in a machine-readable form in the online journal, where also the errors are reported. A part is shown as guidance for the reader regarding its content.}
\label{tab:nxs}
\begin{tabular}{rllccccccccc}
\hline
\hline
ID&\textit{Swift}ID&Counterpart&z&log(N$_H$)&log(\mass)&log(\lx)&log($L_{\rm bol}$)&\eddr&log(\NXS$_{\rm 100s}$)&log(\NXS$_{\rm 1000s}$)\\
& & & & [cm$^{-2}$]&[M$_{\rm sun}$] & [erg\,s$^{-1}$]&[erg\,s$^{-1}$]& & & \\
\hline
\hline
6  & SWIFTJ0006.2+2012  &Mrk\,335    &0.025 & 20.48 &7.23 & 43.23 & 44.36 & 0.068  &-2.30&-2.30 \\  
16 & SWIFTJ0029.2+1319  &PG\,0026+129&0.142 & 20.01 &8.48 & 44.39 & 45.72 & 0.104  &-3.63&-3.03\\   
34 & SWIFTJ0051.6+2928  &UGC\,524    &0.036 & 20.02 &7.62 & 42.99 & 44.08 & 0.032  &-3.20&-3.20 \\ 
36 & SWIFTJ0051.9+1724  &Mrk\,1148   &0.064 & 20.3  &7.75 & 44.12 & 45.31 & 0.234  &-4.54&-5.14 \\ 
39 & SWIFTJ0054.9+2524  &PG\,0052+251&0.155 & 20.02 &8.46 & 44.62 & 45.89 & 0.135  &-3.36&-3.99\\   
43 & SWIFTJ0059.4+3150  &Mrk\,352    &0.014 & 20.01 &7.55 & 42.72 & 44.09 & 0.016  &-3.00&-3.00\\
61 & SWIFTJ0113.8-1450  &Mrk\,1152   &0.052 & 20.01 &8.32 & 43.47 & 45.16 & 0.037  &-3.13&-4.43\\  
73 & SWIFTJ0123.9-5846  &Fairall\,9  &0.047 & 20.02 &8.29 & 44.13 & 45.29 & 0.058  &-2.22&-2.30\\ 
77 & SWIFTJ0127.5+1910  &Mrk\,359    &0.017 & 20.61 &6.04 & 42.66 & 43.83 & 0.339  &-2.69&-2.69 \\ 
106 & SWIFTJ0206.2-0019 &Mrk\,1018   &0.042 & 20.01 &7.81 & 43.61 & 45.08 & 0.094  &-3.61&-4.03\\
...&\\
\hline
\hline 
\end{tabular}\\
{\raggedright Note: The values of the \mass are estimated from RM or broad lines \citep{Koss_2022,2022ApJS..261....5M,2022ApJS..261....8R}. \eddr are computed using the bolometric luminosities calculated from the intrinsic 14--150\,keV luminosities as shown in \citealt{Ricci2017} with a bolometric correction of 8 \citep{2022ApJS..261....1K}. N$_H$ and \lx from \citealt{Ricci2017} \par}
\end{table*}
The \xmm EPIC-pn raw data have been processed using the \textsc{epchain} tool of \textsc{SAS} to obtain calibrated and concatenated event lists. The extraction radii and the optimal time cuts to exclude periods of high flaring particle background were computed via an iterative process which maximizes the signal-to-noise ratio (SNR), as described in \citet{Piconcelli2004}, filtering out those time intervals for which the count rate of the background reach values so high that the SNR of the source does not improve (or even worsens) when including such time intervals in the analysis. Since the pn camera has a full-frame time resolution of $73.3 \rm ms$\,per CCD, the observations generally do not suffer significantly from pile-up, making them suitable for variability analysis. Nonetheless, the light curves were extracted after confirming that the data were not affected by pile-up, as indicated by the SAS task \textsc{epatplot}. The resulting optimal extraction radius was $\sim 30-40$\arcsec\,and the background spectra were extracted from source-free circular regions with radii of $\sim 50-60$\arcsec\, for all the observations analyzed in this work. With these regions we extracted the EPIC-pn source and background light curves using the command \textsc{evselect} and we corrected the source light curve for the background using the command \textsc{epiclccorr}. We extracted the light curves using several different time and spectral binning strategies: 100\,s and 1000\,s in the $0.2-10$\,keV energy band, and 100\,s in the $0.2-1$\,keV (soft), $1-3$\,keV (medium) and $3-10$\,keV (hard) energy bands.\\
Following \citet{Ponti2012} we selected the observations which had cleaned exposure times larger than 10\,ks, and which had at least 10 counts in those 10\,ks chunks and in each (rest-frame) energy band used in this analysis, i.e.  $0.2-1$\,keV, $1-3$\,keV and $3-10$\,keV and for each time bin of 100\,s and 1000\,s. We did this selection to avoid having not enough counts in the 10\,ks independent light curve to constrain the \NXS. A total of 151 sources ($\sim 500$ observations) fulfill these criteria. The distributions of \mass, \lx, \eddr and N$_{\rm H}$ of our sample is shown in Fig.\,\ref{fig:distributions}.\\
We show in Appendix \ref{app:lc} the \xmm EPIC-pn background subctracted light curves of a sub-sample of representative sources for different \mass values (see Fig. \ref{fig:some_lc}).
\section{Analysis}
\label{sect:analysis} 
\begin{figure*}
\includegraphics[width=\textwidth]{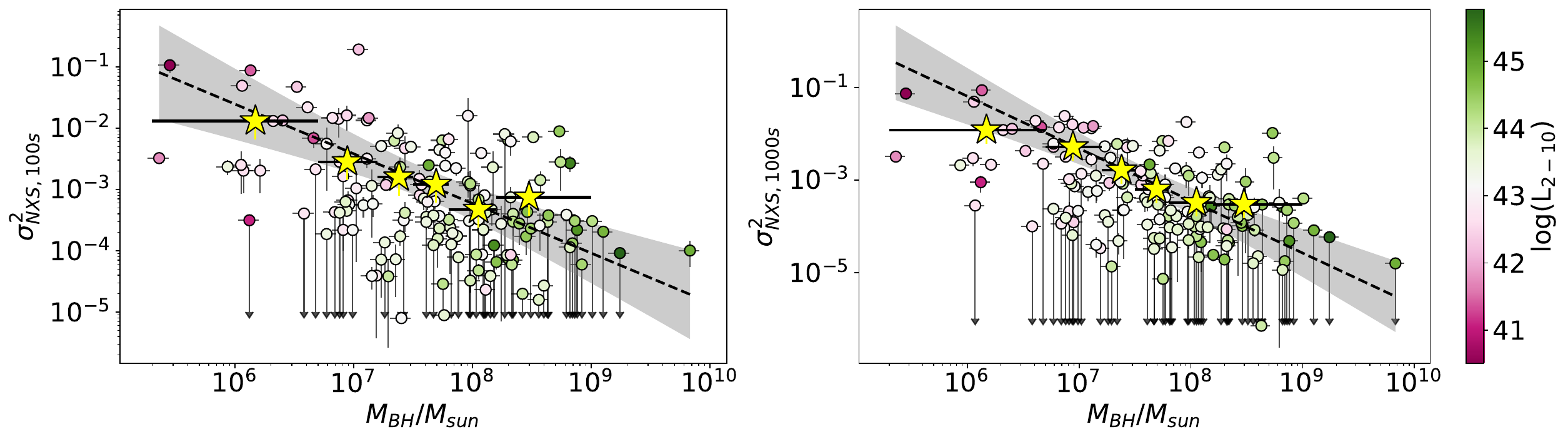}
\includegraphics[width=\textwidth]{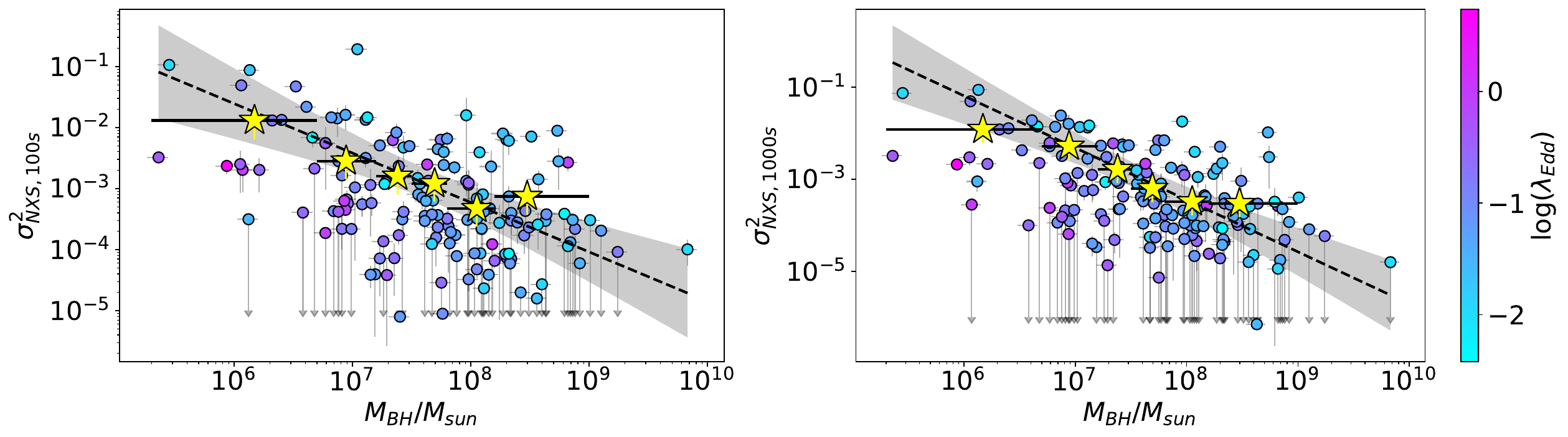}
\caption{\NXS vs \mass relation obtained from the $0.2-10$\,keV light curves binned with 100\,s (left panels) and 1000\,s (right panels). Yellow stars with error bars correspond to the SA results for each bin. The dashed lines are the linear regressions obtained from the fitting over the SA results, while the shaded region represents the combined 1$\sigma$ error on the slope and normalisation. Each colored data points with error bars represents one source of our sample. Colorbars represent the \lx (top panels) and \eddr (bottom panels).}
\label{fig:fit_m_nxs}
\end{figure*}
The excess variance (\NXS) is a quantity used to describe the variability amplitude. It is the difference between the total variance of a light curve and the mean squared error that is normalised for the average of the $N$ flux measurements squared (e.g. \citealt{1997ApJ...476...70N, 1999ApJ...524..667T}). Here $N$ is the number of good time intervals in a light curve, and $x_i$ and $\sigma_i$ are the flux and error in each interval, respectively. The excess variance is defined \citep{vaughan2003} as follows:
\begin{equation}
  \sigma^2_{\rm NXS}=\frac{S^2-\overline{\sigma^2}}{\overline{x_i^2}}
\label{eq:excess_var}
\end{equation}
Where $\overline{\sigma^2}$ is the mean square error:
\begin{equation}
    \overline{\sigma^2}=\frac{1}{N}\sum_{i=1}^{N}[\sigma_i^2]
\end{equation}
and $S^2$ is the sample variance:
\begin{equation}
    S^2=\frac{1}{N-1}\sum_{i=1}^{N}[(x_i-\overline{x_i})^2]
\end{equation}
corresponding to the integral of the PSD between two frequencies ($\nu_1$ and $\nu_2$), which yields the contribution to the expectation value of the variance due to variations between the corresponding timescales ($1/\nu_1$ and $1/\nu_2$): 
\begin{equation}
\langle S^2 \rangle=\int_{\nu_1}^{\nu_2}P(\nu)d\nu
\end{equation}
To study the correlations between \NXS and \mass, \lx and \eddr, we calculated \NXS from the \xmm light curves. The values are listed in Tab.\,\ref{tab:nxs}.\\
\NXS is a good estimator of the intrinsic variance of a source but it has some biases. It is related to the integral of the PSD between two frequencies and thus depends on the length of the monitoring time interval, on the red-noise character of the X-ray variability and also, due to the effect of cosmological time dilation, on the redshift \citep{1993ApJ...414L..85L,1993MNRAS.265..664G,1993ApJ...414L..85L,2008A&A...487..475P,2011A&A...536A..84V,Vagnetti2016}. Since our sample of 151 type\,1 AGNs is composed mainly of local AGNs ($z_{\rm med}=0.035$), the impact of redshift is negligible. However, we need to avoid biases related to the different exposure times of our observations and the red-noise character of the light curves. Therefore, we computed the \NXS from 10\,ks-long independent light curve sections and, for the sources with cleaned exposure time that lasted for a multiple of 10\,ks, we took the median of the excess variances of all these independent sections in each energy band. For the sources with more than one observation, we used the median value of the \NXS in both the cases of the light curves with 100\,s and 1000\,s time bin in each energy band. We applied this procedure also for the 7 sources of our sample which are classified as 'changing-look' (CL) AGNs (i.e. Mrk\,1018, Fairall\,9, Mrk\,590, NGC\,3516, NGC\,1566, 3C\,390.3, NGC\,7603, \citealp{2022ApJ...926..184J,2023MNRAS.518.2938T}) since the \NXS computed for these sources are consistent within the error among the different observations.\\
The X-ray spectrum of AGNs in different energy bands is strongly impacted by different components: the primary power-law component and the reflection component are dominant in the hard energy band ($3-10$\,keV, \citealt{1991ApJ...380L..51H,1993ApJ...413..507H,1993MNRAS.261..346H}) while soft-excess and warm-absorbers (WA) can impact the soft ($0.2-1$\,keV, \citealt{Bianchi2009}) and medium ($1-3$\,keV, \citealt{Blustin2005,Tombesi2013}) energy bands. Variations of these different components will lead to distinct spectral variability in different energy bands. We therefore calculated \NXS from the $0.2-1$\,keV (soft), $1-3$\,keV (medium) and $3-10$\,keV (hard) light curves to get a fuller picture of the AGNs X-ray variability. 
\begin{figure*}
\includegraphics[width=\textwidth]{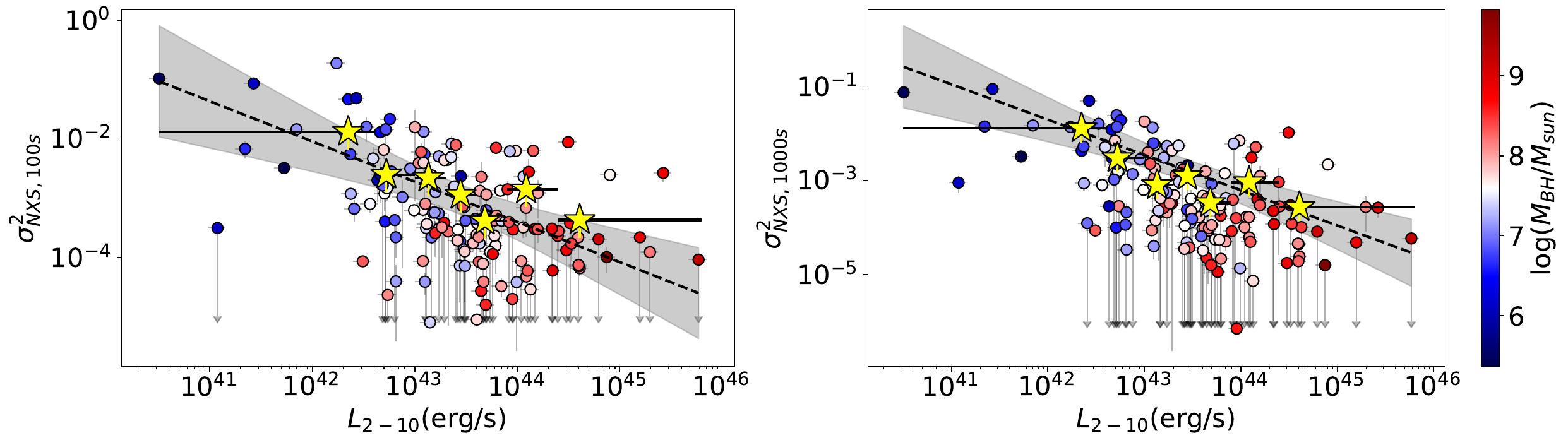}
\includegraphics[width=\textwidth]{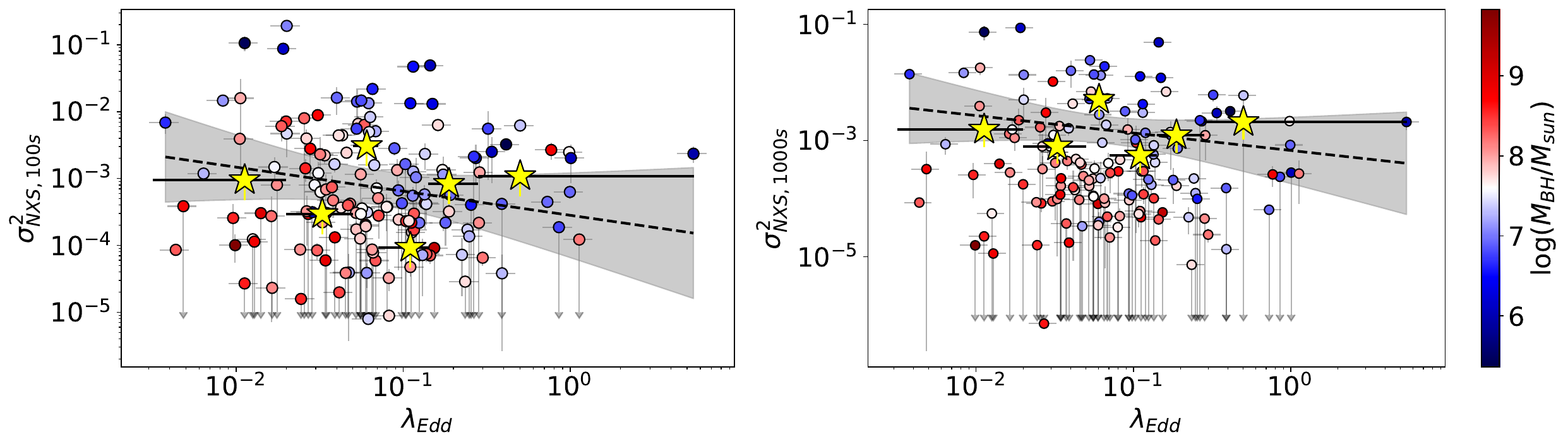}
\caption{Upper panels: \lx vs \mass relation. Lower panels: \eddr vs \NXS relations. The relations are obtained from the $0.2-10$\,keV light curves binned with 100\,s (left panels) and 1000\,s (right panels). Yellow stars with error bars correspond to the SA results for each bin. The dashed lines are the linear regressions obtained from the fitting over the SA results, while the shaded region represents the combined 1$\sigma$ error on the slope and normalisation. Each colored data points with error bars represents one source of our sample. Colorbars represent the \mass.}
\label{fig:fit_l-edd_nxs}
\end{figure*}
\begin{figure*}
\includegraphics[width=\textwidth]{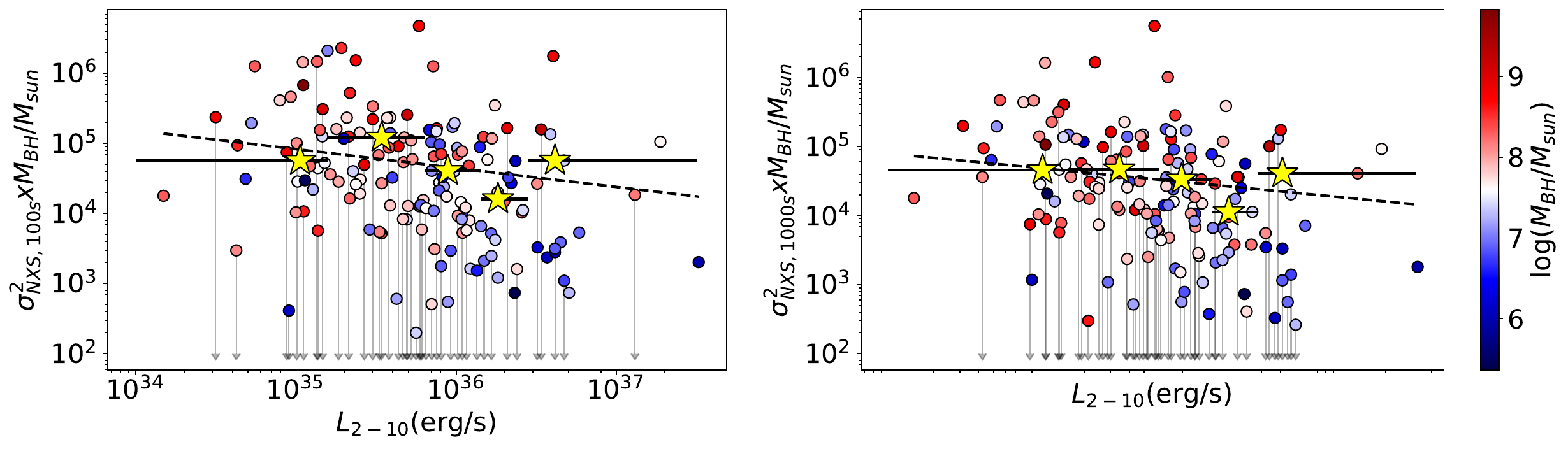}
\caption{\NXS x \mass vs \lx relation. The relations are obtained from the $0.2-10$\,keV light curves binned with 100\,s (left panels) and 1000\,s (right panels). Yellow stars with error bars correspond to the SA results for each bin. The dashed lines are the linear regressions obtained from the fitting over the SA results. Colorbars represent the \mass.}
\label{fig:nxspm}
\end{figure*}
\begin{figure*}
\centering
\includegraphics[width=\textwidth]{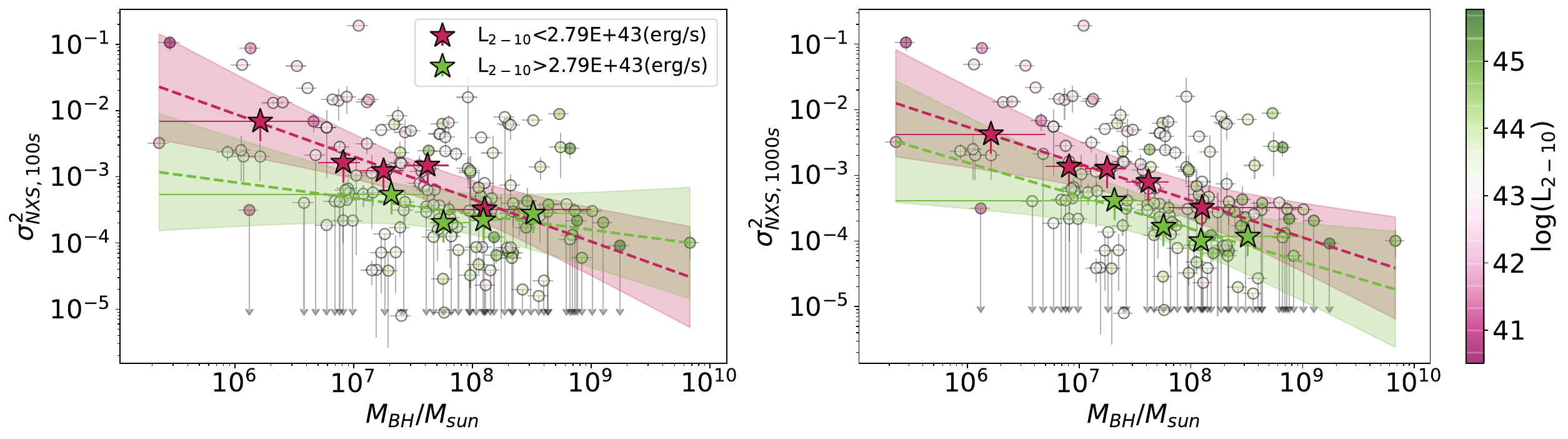}
\includegraphics[width=\textwidth]{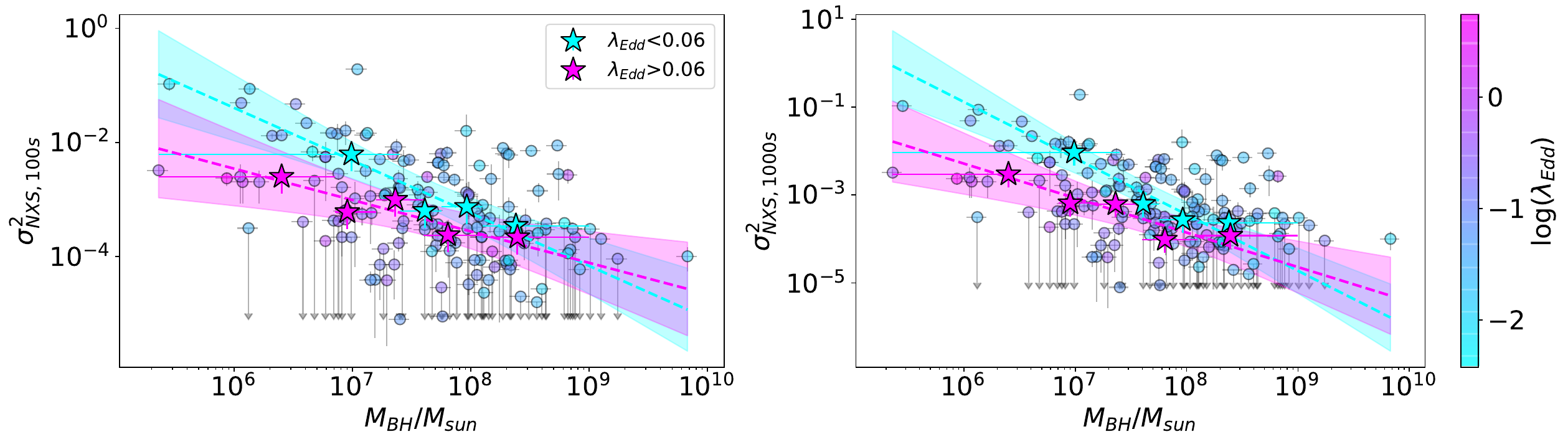}
\caption{Relations between the \NXS vs \mass when the sample is divided into two bins of \lx (top panels, purple lines: $L_{\rm 2-10}<L_{\rm 2-10, med}$, green lines: $L_{\rm 2-10}>L_{\rm 2-10, med}$) and \eddr (bottom panels, cyan lines: $\lambda_{\rm Edd}<\lambda_{\rm Edd, med}$, magenta lines: $\lambda_{\rm Edd}>\lambda_{\rm Edd, med}$). The best fit relation are reported in Tab.\,\ref{tab:relations}.}
\label{fig:two_bins}
\end{figure*}
\begin{table*}
\caption{List of the best-fit relations together with their p-values. The fits are performed in the log-log space using equation Eq. \ref{eq:fitting_rel}. }
\label{tab:relations}
\begin{tabular}{lccccc}
\hline 
\hline
Relation&$\Delta$t(s)&Intercept (A) &Slope (B) &Pearson&1-P$_{\rm value}$\\
\hline
\hline
\noalign{\medskip}
\NXS vs \mass&100&$2.99 \pm 1.79$& $-0.77 \pm 0.23$&-0.94&0.99\\
\NXS vs \mass&1000& $5.54 \pm 2.11$&$-1.12 \pm 0.18$&-0.96&0.99\\
\noalign{\medskip}
\NXS vs \lx & 100&$26.29 \pm 25.29 $&$-0.67 \pm 0.08$&-0.84&0.95\\
\NXS vs \lx & 1000&$29.84 \pm 36.41$&$-0.76 \pm 0.06$&-0.85&0.96\\
\noalign{\medskip}
\NXS vs \eddr & 100&$-3.55\pm 2.27$ & $-0.36\pm 0.61$&-0.10&0.25\\
\NXS vs \eddr & 1000&$-3.11 \pm 3.19$& $-0.24\pm 0.59$&-0.12&0.25\\
\noalign{\medskip}
\hline
\noalign{\medskip}
\NXS vs \mass$_{\rm, RM}$&100&$2.94\pm0.29$& $-0.75 \pm 0.20$&-0.66&0.75\\
\NXS vs \mass$_{\rm, RM}$&1000& $2.91 \pm 0.34$&$-0.76 \pm 0.41$&-0.67&0.79\\
\noalign{\medskip}
\hline
\noalign{\medskip}
\NXS vs \mass\,(\lx$<L_{\rm 2-10, med})$&100 &$0.36\pm0.51$&$-0.43\pm0.06$&-0.89&0.97\\
\NXS vs \mass (\lx$<L_{\rm 2-10, med})$&1000 &$0.16\pm0.38$&$-0.37\pm0.04$&-0.99&0.99\\
\noalign{\medskip}
\NXS vs \mass (\lx$>L_{\rm 2-10, med})$&100 &$-2.45\pm0.48$&$-0.14\pm0.16$&-0.30&0.38\\
\NXS vs \mass(\lx$>L_{\rm 2-10, med})$&1000 &$-1.71\pm0.41$&$-0.16\pm0.15$&-0.35&0.41\\
\noalign{\medskip}
\hline
\noalign{\medskip}
\NXS vs \mass(\eddr$<\lambda_{\rm Edd, med})$&100 &$0.69\pm0.76$&$-0.48\pm0.09$&-0.89&0.97\\
\NXS vs \mass(\eddr$<\lambda_{\rm Edd, med})$&1000 &$1.46\pm1.32$&$-0.55\pm0.16$&-0.89&0.97\\
\noalign{\medskip}
\NXS vs \mass(\eddr$>\lambda_{\rm Edd, med})$&100 &$0.16\pm0.83$&$-0.42\pm0.13$&-0.89&0.97\\
\NXS vs \mass(\eddr$>\lambda_{\rm Edd, med})$&1000 &$2.35\pm0.88$&$-0.71\pm0.11$&-0.89&0.97\\
\noalign{\medskip}
\hline
\noalign{\medskip}
$\sigma^2_{\rm NXS} \times M_{\rm BH} $ vs $L_{\rm 2-10}$& 100 & $14.35 \pm 9.75$ & $-0.26 \pm 0.27$ & -0.19 & 0.30\\
$\sigma^2_{\rm NXS} \times M_{\rm BH} $ vs $L_{\rm 2-10}$& 1000& $12.05\pm 7.78$ & $-0.21 \pm 0.22$ & -0.20 & 0.30\\
\noalign{\medskip}
\hline
\noalign{\medskip}
 $\sigma^2_{\rm NXS, hard}$ vs $\sigma^2_{\rm NXS, soft}$ &100&$0.62\pm0.26$& $0.87\pm0.10$&0.71&0.99\\
$\sigma^2_{\rm NXS, hard}$ vs $\sigma^2_{\rm NXS, med}$& 100&$0.47 \pm 0.22$&$0.93 \pm 0.08$&0.79&0.99\\
\noalign{\medskip}
\hline
\hline
\end{tabular}
\end{table*}
\subsection{Correlations between the normalised excess variance and the physical parameters}
To investigate the physical parameters driving X-ray variability in our sample of unobscured AGNs we looked for correlations between the \xmm broad-band ($0.2-10$\,keV) \NXS and several key AGNs parameters (i.e. \mass, \lx, \eddr) by fitting a linear model to the data in the log-log space (see Fig.\,\ref{fig:fit_m_nxs} and Fig.\,\ref{fig:fit_l-edd_nxs}) using the following fitting relation:
\begin{equation}
    \log(\sigma_{\rm NXS}^2)=\rm{A}+\rm{B}\log(x)
    \label{eq:fitting_rel}
\end{equation}
where x is the value of the physical parameter.
Among the 151 sources of our sample, we found 46 objects with an intrinsic \NXS lower than the respective error. In this case, we define the measurement as a “non-detection”, and we consider it as an upper limit. To include the upper limits in our analysis, we used the survival analysis method (SA; e.g., \citealt{1985ApJ...293..192F,2017MNRAS.466.3161S}) using the \textsc{scikit-survival} \citep{sksurv} package, which applies the principles of SA to astronomical data. SA is a statistical technique used to analyze time-to-event data and it is particularly well-suited for analyzing data that include upper/lower limits. Specifically, \textsc{scikit-survival} calculates the non-parametric Kaplan-Meier product-limit (KMPL) estimator for a sample distribution. The KMPL estimator is an estimate of the survival function, which is simply 1-CDF (cumulative distribution function). Using the KMPL, we calculated, for each bin of \mass, \lx and \eddr, the median \NXS, and estimated their uncertainties. Since the KMPL estimator is a non-parametric method, it is unbiased because it does not assume any specific distribution for the data. We divided \mass and \eddr into 6 bins and the \lx into 7 bins. These bins are not symmetrical, since we requested each bin to have at least 15 values. We fitted the median values obtained with the SA method using the code: \textsc{linmix}, a hierarchical Bayesian model for fitting a straight line to data with errors in both the x and y directions \citep{Kelly_2007}. From the analysis of the correlation between \NXS and \mass, data of sources with \mass$>10^9M_{\odot}$ are excluded since, for these sources, we found mostly upper limits on \NXS so, being the last bin populated just by upper limits, the SA method was not reliable in computing the median in the bin.\\ 
We report in Tab.\,\ref{tab:relations} the intercepts and the slopes of the linear regression line together with Pearson’s correlation coefficients and the correlation probabilities for all the relations analysed in this work and for light curves with 100\,s and 1000\,s bins. As shown in Tab.\,\ref{tab:relations} the fitting parameters for the 100\,s and 1000\,s binned light curves are consistent within the errors in each analysed relation.\\
Fig.\,\ref{fig:fit_m_nxs} shows the \NXS vs \mass relation obtained from the $0.2-10$\,keV light curves binned with 100\,s and 1000\,s. We also reported the SA results for each bin. Being the SA results the \NXS median calculated using the KMPL estimator they are representative of the \NXS value in each \mass bin. We report also the linear regressions obtained from the fitting over the SA results. As expected from previous studies \citep{1997ApJ...476...70N,2004MNRAS.348..207P,2005MNRAS.358.1405O,Ponti2012}, we found, a strong anti-correlation between \NXS and \mass (see Fig.\,\ref{fig:fit_m_nxs}). We computed the 1$\sigma$ scatter of the data around the best-fit line using the following equation:
\begin{equation}
    \sigma_{\rm scatter}= \sqrt{\sum_{i=1}^N[\log(\sigma^2_{\rm NXS,i})-f(M_{\rm BH,i})]^2/N}
\label{eq:scatter}
\end{equation}
where $f(M_{\rm BH})$ is the logarithmic value of the \NXS extrapolated using the best fitting relation (see Tab.\,\ref{tab:relations}).
We found a scatter for this relation of $\sim0.85$\,dex in both the case of the light curves binned with 100\,s and 1000\,s.\\
We also found a strong anti-correlation between the \NXS and the $L_{\rm 2-10}$ (see upper panels of Fig.\,\ref{fig:fit_l-edd_nxs}). Given the strong dependence between the \NXS and \mass, we decided to correct the \NXS for \mass to check if after the correction, the relation between \NXS and \lx is still present. Previous works have shown that \NXS $\propto$ \mass$^{\sim(-1)}$ (e.g., \citealt{Ponti2012}). In this work we found a slope for the \NXS vs \mass relation of $-0.77\pm0.23$ and $-1.12 \pm 0.18$ when considering light curves with 100\,s and 1000\,s bins, respectively. Since $-1$ is still consistent within the errors with our results, for consistency with past works, to check if the \NXS vs \lx correlation still exists when the primary dependence is removed, we analysed the correlation between $\sigma^2_{\rm NXS} \times M_{\rm BH}$ versus \lx. Removing the dependence of \NXS on \mass, the strong correlation with \lx, that was present before, is not significant anymore (see Fig.,\ref{fig:nxspm}), as reported from previous studies \citep{2004MNRAS.348..207P,2005MNRAS.358.1405O}.
In fact we found, in both cases of \NXS obtained from the $0.2-10$\,keV light curves binned with 100\,s and 1000\,s, a Pearson correlation coefficient of -0.19 and -0.20 corresponding to a $1-P_{\rm value}$ of 0.30 (see Tab.\,\ref{tab:relations}). Thus, the dependence between the \NXS and $L_{\rm 2-10\,keV}$ is actually related to the dependence between $L_{\rm 2-10\,keV}$ and \mass. From our analysis an anti-correlation between \NXS and \eddr is present (see lower panels of Fig.\,\ref{fig:fit_l-edd_nxs}), but it is not significant, according to the Pearson test (see Tab.\,\ref{tab:relations}).\\
From Fig.\,\ref{fig:fit_l-edd_nxs} it is clear that in both relations a gradient of \mass is present. Thus, to check if the relation $\sigma_{\rm NXS}^2 - M_{\rm BH}$ is somehow affected by \lx and/or \eddr, we first computed the median values of \lx and \eddr of the sample, which are $L_{\rm 2-10, med}=2.79\times10^{43}$\, $\rm erg\,s^{-1}$ and $\lambda_{\rm Edd, med}=0.06$, respectively. We then used these values as thresholds to divide the sample into two sub-samples depending on their \lx and \eddr (see figure \ref{fig:two_bins}). 
The best-fitting values of the correlations we found are also reported in Tab.\,\ref{tab:relations}. We found that the correlation between \NXS and \mass has slightly different normalizations among the two sub-samples depending on the \lx or \eddr but the same slope within the errors, which is also in agreement with the slope of the \NXS - \mass relation found for the total sample. In particular, in the sub-sample depending on \lx we found that the slopes of the correlations is similar to the one found for the total sample, while the normalization for the sources with $L_{2-10}>L_{\rm 2-10, med}$ is lower. This is not really surprising since sources with high luminosity (high \mass) show lower variability. Also, the majority of the sources with $L_{2-10}>L_{\rm 2-10, med}$ show an upper limit of the \NXS.
\subsection{Reverberation mapping sub-sample}
\label{sect:RM}
\begin{figure*}
\includegraphics[width=\textwidth]{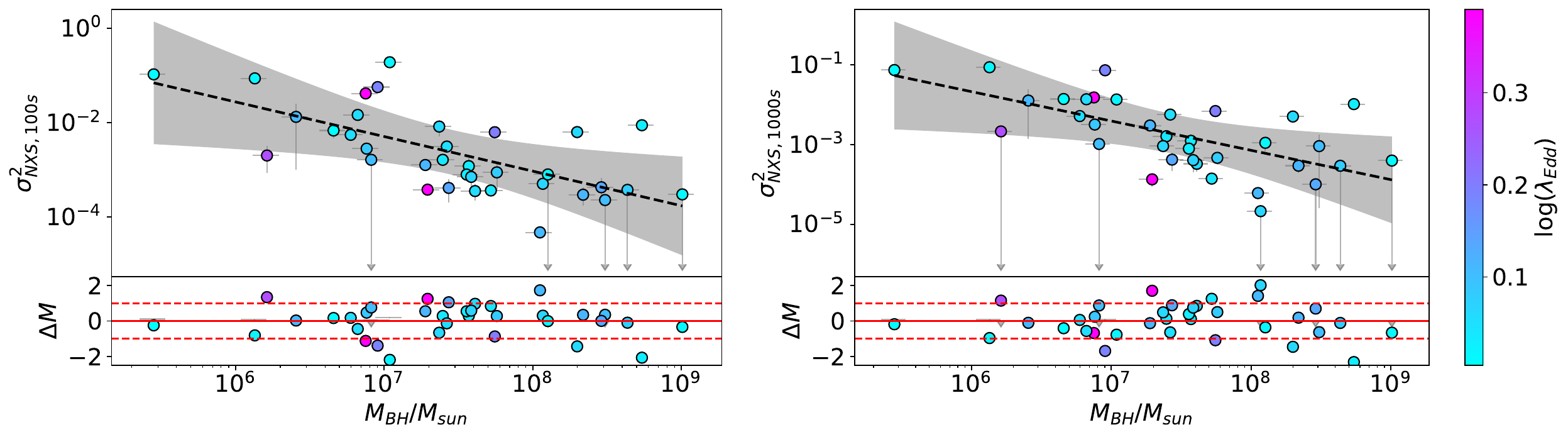}
\caption{\NXS vs \mass for the RM sub-sample. The dashed black lines are the linear regressions obtained from the fitting process. Shaded region represents the combined 1$\sigma$ error on the slope and normalisation. The \NXS is obtained from the $0.2-10$\,keV light curves binned with 100\,s (left panel) and 1000\,s (right panels). Lower panels show the difference between the real \mass and \mass extrapolated from the relation ($\Delta M$).
For details about the coefficients, see Tab.\,\ref{tab:relations}.}
\label{fig:RM}
\end{figure*}
We checked the relation between the \NXS and \mass in the sub-sample of sources in which \mass is obtained via RM (35 sources). Being the sub-sample smaller, we did not used the SA method. Instead we used the method of the "censored fitting" (CF) \citep{Guainazzi2006,Bianchi2009}, to account for upper limits. This was done by performing a large number of least square fits, using the \textsc{linmix} code, on a set of Monte-Carlo simulated data derived from the observed data points. Each detection was substituted by a value randomly drawn from a Gaussian distribution, whose mean is the best-fit measurement and whose standard deviation is its statistical uncertainty. Each upper limit $U$ was substituted by a value randomly drawn from a uniform distribution in the interval $[A,U]$, where A was arbitrarily set to $A \ll U$. We choose $A=10^{-6}$. We found an anticorrelation in both cases of \NXS obtained from the $0.2-10$\,keV light curves binned with 100\,s and 1000\,s (see Fig.\,\ref{fig:RM}). 
Using Equation\,\ref{eq:scatter} we found $\sigma_{\rm scatter,100}= 0.65$ and $\sigma_{\rm scatter,1000}= 0.69$, smaller than the scatters of the \NXS vs \mass relations obtained from the total sample. This is because on average the sources with \mass estimated via RM are brighter and they show a higher count rate on the same timescales. Using the \NXS vs \mass relations obtained from the RM sample, it is possible to measure \mass for a total of 87 AGNs (out of 151 AGNs in our sample) and provide an upper/lower limit for the remaining AGN. Thus, even if the X-ray variability is not the most accurate tool to measure \mass, the relation obtained for the RM sub-sample gives a good \mass estimation, with a scatter <1\,dex. 
\subsection{The normalised excess variance in the soft, medium and hard energy bands.}
\label{sect:soft-med-hard}
\begin{figure*}
\includegraphics[width=\textwidth]{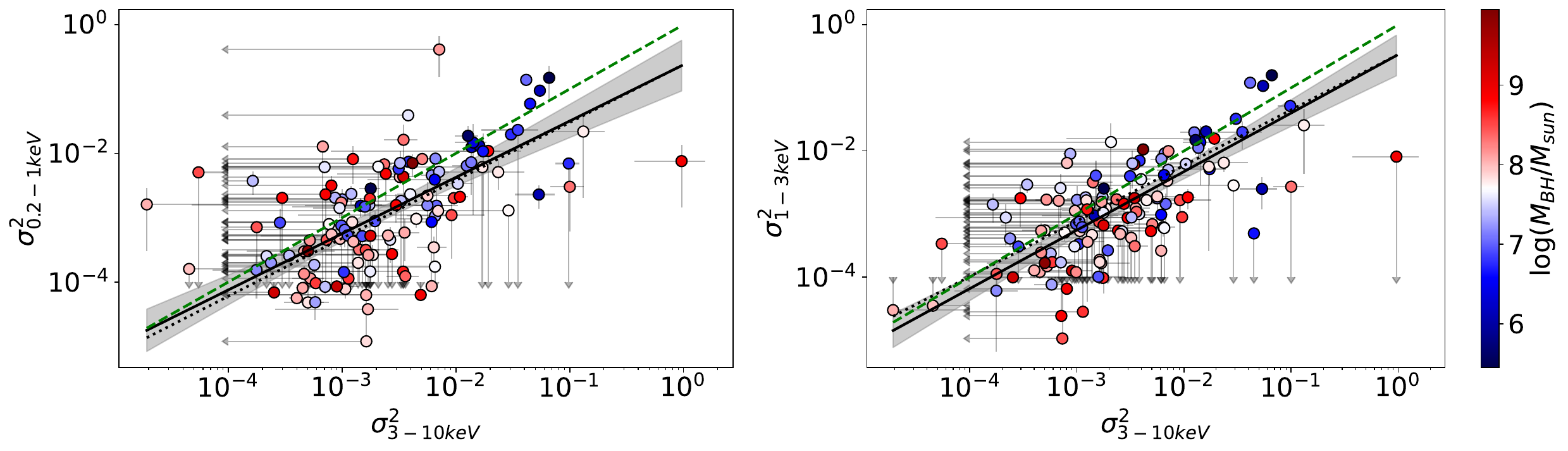}
 \caption{Left panel: soft ($0.2-1$\,keV) vs hard ($3-10$\,keV) \NXS. Right panel: medium ($1-3$\,keV) vs hard ($3-10$\,keV) \NXS. The best fit curves are plotted with solid black lines while the shaded region represents the combined 3$\sigma$ error on the slope and intercept. The green dashed lines represent the one-to-one relation. Black dashed lines represent the regression lines found using the bisector method.} 
 \label{fig:sigma_soft_med_hard}
\end{figure*}
It is interesting to compare \NXS in various energy bands to check if there is one or more components of the AGNs X-ray spectrum that contributes the most in the variability. In order to verify this we calculated \NXS in the the soft ($0.2-1$\,keV), medium ($1-3$\,keV) and hard ($3-10$\,keV) bands. We then looked for a correlation between the $\sigma^2_{\rm NXS, hard}$ and both $\sigma^2_{\rm NXS, soft}$ and $\sigma^2_{\rm NXS, med}$. In the left panel of Fig.\,\ref{fig:sigma_soft_med_hard} we show \NXS in the soft energy band versus the same parameter calculated in the hard band, while in the right panel of Fig.\,\ref{fig:sigma_soft_med_hard} we illustrate the excess variance in the medium energy band versus that in the hard band. The best-fitting relations, reported in Tab.\,\ref{tab:relations}, are obtained using the CF method (see \S \ref{sect:RM}).\\
We fitted the data also with the bisector method which provides a more balanced and symmetric estimate of the true regression line between the two variables that in principle could be independent. The values of the slope and intercept we found with this method are consistent within the errors with the one found using the CF method.\\
The values of \NXS in the soft and medium energy bands appear to be well correlated with \NXS in the hard energy band, although with a slope flatter than the one-to-one relation (green dashed line in Fig.\,\ref{fig:sigma_soft_med_hard}).
This may imply that for most of our sources, on timescales less than 10\,ks the spectral components which dominate in the hard band are increasingly more variable than the ones dominating the soft and medium energy bands, for higher values of the variance. In the soft energy band the dominant component is usually the soft-excess, which can be variable. Furthermore, in the soft-medium energy bands, the presence of absorbing material, either neutral or ionised, can be variable or can absorb the continuum emission responding to the continuum variations. If these variations happened on timescales less than $\sim 10$\,ks we would expect to measure a larger variability amplitude in the soft/medium energy bands compared to the hard band. We observe the opposite, i.e. weaker variations of the soft-excess and/or warm absorbers than those of the primary continuum and/or reflection component on this timescale, in agreement with previous studies \citep{Ponti2012,Simm2016}.\\
For completeness we checked the relations between the \NXS in the soft, medium and hard energy bands with the physical properties of the AGN (\mass, \lx and \eddr) to see whether these relations support the results found in this work. The best fitting results are reported in Appendix \ref{app:smh}. We found that the \NXS vs \mass relation in the soft energy band is slightly less significant than in the harder energy bands (see Tab. \ref{tab:relations_smh}), but in general the best-fitting relations in each energy band are consistent with those from the broad \xmm energy band ($0.2-10$\,keV).
\section{Conclusions}
\label{sect:conclusion}
We analysed the variability properties of $\sim 500$ \xmm observations of a sample of 151 nearby ($z_{\rm med}=0.035$) unobscured ($N_{H}<10^{22}$cm$^{\rm-2}$) AGNs from the BASS survey, studying the correlations of the excess variance with the physical properties of the sources and also checking for the correlations between the excess variance computed in different energy bands. The timescale used to compute the \NXS is 10\,ks, to avoid biases related to the differences on the exposure times of the sources of our sample and to take into account the red-noise character of the light curves. We analysed the relations of \NXS with \mass, \lx and \eddr. The correlation between \NXS and \mass is a well-know property of AGNs \citep{2001MNRAS.324..653L,2004MNRAS.348..207P,2005MNRAS.358.1405O,2006MNRAS.370.1534N,2007ASPC..373...66N,2009MNRAS.394..443M,2010ApJ...710...16Z,Ponti2012}. In agreement with this, in our sample we found a very strong and highly significant correlation between these two quantities.\\
We do not find a significant correlation between \NXS and \eddr, consistently with previous results \citep{2005MNRAS.358.1405O,2008MNRAS.383..741G,2010ApJ...710...16Z,Ponti2012,Lanzuisi2014}.
However, according to \citet{McHardy2006} the break timescale increases proportionally as \mass decreases (\eddr increases). Thus, if we assume a universal PDS with a single break frequency depending on the \mass and a equally long-timescale normalization, the strength of the relations \NXS vs \mass and \NXS vs \eddr would be the same, with higher short-timescale variability for low \mass (high \eddr). Our result could suggest that there might be no correlation between break timescale and \eddr, as proposed by \citet{2012A&A...544A..80G} analysing a larger sample of shorter light curves. Alternatively, following the results of \citet{McHardy2006} and \citet{2017MNRAS.471.4398P}, \eddr could be dependent on the break time scale but, since on short-timescale \NXS seems mostly independent of this parameter, the normalization of the power spectrum may be anti-correlated with \eddr.\\
We found a tight anti-correlation between \NXS and \lx. To remove the \mass dependence from this correlation, we explored the relation between the $\sigma^2_{\rm NXS} \times M_{\rm BH}$ versus \lx, finding that, in this case, the \NXS vs \lx correlation disappears, confirming that the correlation with \lx is secondary, while the primary correlation is in fact with the mass, in agreement with what has been found by previous works (e.g. \citealt{2004MNRAS.348..207P,2005MNRAS.358.1405O,Lanzuisi2014}).\\ 
We explored the \NXS vs \mass relation in the sub-sample of sources with \mass estimated via RM, finding that the correlation between these quantities in this sub-sample has an intrinsic scatter of $\sim 0.65-0.69$\,dex. With this relation we were able to measure \mass for 87 AGNs and estimate upper/lower limits for the remaining 64 AGNs of our sample. Thus, one could in principle use X-ray variability to measure \mass (e.g. \citealt{10.1111/j.1365-2966.2004.07829.x,Ponti2012,2022A&A...666A.127A}). With the advent of future planned or proposed missions (e.g. \textit{Athena}, \textit{AXIS}, etc) that will provide higher count rates, the accuracy of this relation for mass measurement will improve significantly.\\
Dividing the sample into two bins of \lx, the normalization of the anti-correlation between \NXS and \mass is lower for the sources with higher luminosity. This is possibly related to the fact that for sources with high luminosity (high mass), we detected lower variability and mostly only an upper limit on \NXS was obtained. When dividing the sample into two bins of \eddr the slope and the normalisation are slightly different in the two sub-samples but still consistent within the errors.\\ 
X-ray spectra could be dominated by different components depending on the energy band one is analysing. Therefore we explored the relation of the \NXS in the hard X-ray band ($3-10$\,keV) with the \NXS in the soft X-ray band ($0.2-1$\,keV) and in the medium X-ray band ($1-3$\,keV), finding that \NXS calculated in various energy bands are highly correlated, in agreement with previous studies \citep{Ponti2012,Simm2016}. In particular we found that, in most sources, the primary continuum and/or the reflection component are increasingly more variable than the spectral components dominating softer energy bands ($0.2-1$\,keV and $1-3$\,keV) on timescales shorter than 10\,ks. In fact, if WA components were varying, they would show more variability in the medium energy band, while the variance in that band is lower than in the hard energy band. Thus WA variability cannot be generally the cause of fast (shorter than 10\,ks) variations. Moreover, we found that the soft energy band is less variable than the hard band. This implies that the soft-excess, or at least part of it, is a less variable component (on timescales less than 10\,ks) which dilutes the \NXS by adding to the constant flux in the denominator and not to the variable flux in the numerator, as it was found for the Seyfert 1.5 galaxy NGC\,3227 \citep{2014ApJ...783...82A}. Finally, the hard continuum might be intrinsically more variable than the continuum in softer bands because the break timescale of the PDS moves to shorter timescales for higher energy X-ray photons \citep{2004MNRAS.348..783M,2007ApJ...656..116M,2007MNRAS.382..985M,2008MNRAS.388..211A}.\\
We examined the relation between \NXS (calculated in the soft, medium and in the hard X-ray bands) and several important AGN physical parameters, such as \mass, \lx and \eddr. Our analysis revealed that the best-fitting relations in each energy band align with those from the broad \xmm energy band ($0.2-10$\,keV). Notably, in the soft energy band, the \NXS vs \mass anti-correlation appears to be slightly less significant. This lends support to another key finding of this study, i.e. that, on timescales shorter than 10\,ks, the primary continuum and/or the reflection component exhibit stronger variability compared to the spectral components dominating softer energy bands.\\
Compared with previous results (e.g., \citealt{Ponti2012}) we found a less steep correlation between \NXS and \mass. The difference could be attributed to the larger number of black hole masses from reverberation mapping, to the higher quality optical measurements and fitting for the other mass measurements techniques, and also to the larger number of observations used (a factor $\sim2$ larger than \citealp{Ponti2012}), which helped to refine the computation of \NXS. Our results are consistent with the common picture in which, as a general rule, nearby AGNs display similar patterns of variability once they are rescaled for \mass and \eddr.
\section*{Acknowledgments}
This work was funded by ANID programs FONDECYT Postdoctorado - 3190213 (AT), 3220516 (MT), 3210157(ARL); FONDECYT Regular - 1230345 (CR) and1200495 (FEB); Millennium Science Initiative Program - ICN12\_009 (FEB); CATA-BASAL - ACE210002 (FEB) and FB210003 (CR, FEB). TL acknowledges support from the NANOGrav NSF Physics Frontiers Center No. 2020265. BT acknowledges support from the European Research Council (ERC) under the European Union's Horizon 2020 research and innovation program (grant agreement number 950533) and from the Israel Science Foundation (grant number 1849/19). TK is supported by JSPS KAKENHI grant No. 23K13153 and acknowledges support by the Special Postdoctoral Researchers Program at RIKEN. KO acknowledges support from the National Research Foundation of Korea (NRF-2020R1C1C1005462) and the Korea Astronomy and Space Science Institute under the R\&D program(Project No. 2023-1-868-03) supervised by the Ministry of Science and ICT. This work is based on observations obtained with the ESA science mission \xmm, with instruments and contributions directly funded by ESA Member States and the USA (NASA). The authors thank the anonymous referee for constructive comments that have helped in improving the quality of the paper . 
\section*{Data Availability}
All the data utilized in this paper are publicly available in the \xmm data archive at \url{https://nxsa.esac.esa.int/nxsa-web/#search}. More details of the observations are listed in Tab.\,\ref{tab:obsid}.

\bibliographystyle{mnras}
\bibliography{bibliography} 

\appendix
\section{Light Curves}
\label{app:lc}
\begin{figure*}
	\includegraphics[width=\textwidth]{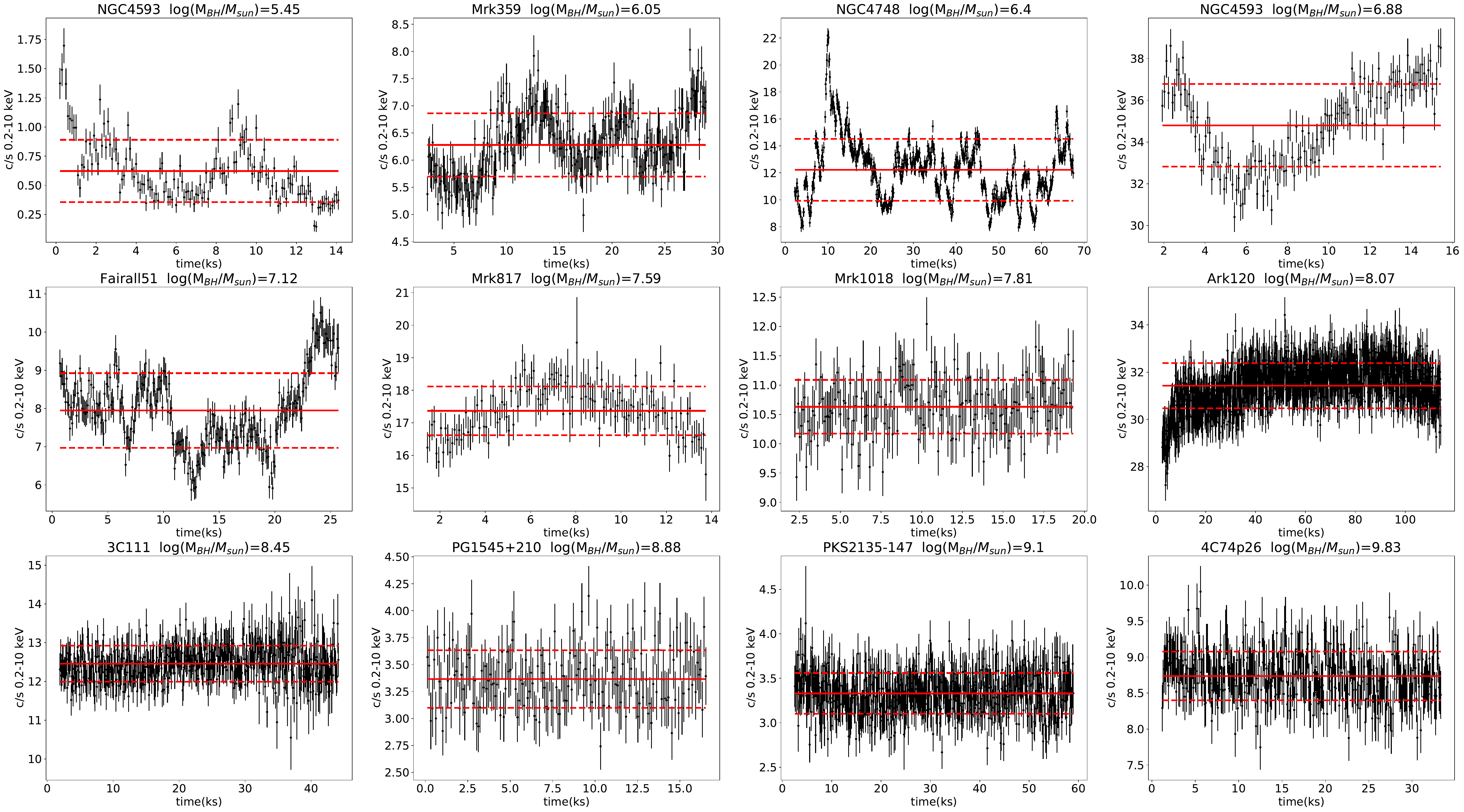}
\caption{\xmm EPIC-pn light curves (background subtracted) in the $0.2-10$\,keV with a time binning of 100\,s are shown for some representative sources of the sample. From the top to the bottom and from the left to the right we reported the light curves for: NGC\,4593, Mrk\,359, NGC\,4748, NGC\,4593, Fairall\,51, Mrk\,817, Mrk\,1018, Ark\,120, 3C\,111, PG\,1545+210, PKS\,2135$-$147, 4C74+26. The red solid and dashed lines indicate the average and the standard error of the mean, respectively.}
\label{fig:some_lc}
\end{figure*}
We show in Fig. \ref{app:lc} the \xmm EPIC-pn light curves (background subtracted) in the $0.2-10$\,keV energy band and with a time binning of 100\,s for some representative sources of the sample: 
\begin{itemize}
    \item NGC\,4593, $\log(M_{\rm BH}/M_{\odot})=4.45$;
    \item Mrk\,359, $\log(M_{\rm BH}/M_{\odot})=6.05$;
    \item NGC\,4748, $\log(M_{\rm BH}/M_{\odot})=6.40$;
    \item NGC\,4593, $\log(M_{\rm BH}/M_{\odot})=6.88$;
    \item Fairall\,51, $\log(M_{\rm BH}/M_{\odot})=7.12$;
    \item Mrk\,817, $\log(M_{\rm BH}/M_{\odot})=7.59$;
    \item Mrk\,1018, $\log(M_{\rm BH}/M_{\odot})=7.81$;
    \item Ark\,120, $\log(M_{\rm BH}/M_{\odot})=8.07$;
    \item 3C111 $\log(M_{\rm BH}/M_{\odot})=8.45$;
    \item PG\,1545+210, $\log(M_{\rm BH}/M_{\odot})=8.84$;
    \item PKS\,2135-147, $\log(M_{\rm BH}/M_{\odot})=9.10$;
    \item 4C74p26, $\log(M_{\rm BH}/M_{\odot})=9.83$.
\end{itemize}

\section{Relations between the excess variance in the soft, medium and hard energy band and the AGN physical quantities}
\label{app:smh}
We checked the relations between \NXS in the soft ($0.2-1$\,keV), medium ($1-3$\,keV) and hard ($3-10$\,keV) energy bands with \mass, \lx and \eddr to see whether these relations support the results found in this work or not. For the analysis we applied the same method described in \S \ref{sect:analysis}. The best fitting results are shown in Tab. \ref{tab:relations_smh}. For both \NXS vs \mass and \NXS vs \lx, the relations we found in the different energy bands are consistent with the ones in the total \xmm energy band (0.2-10\,keV). It is worthwhile to underline that, in the case of the \NXS vs \mass relation, the anti-correlation is slightly less significant in the soft energy band. This can be related to the other result of this paper according to which the primary continuum and/or of the reflection component (the spectral component dominating the hard energy band) are increasingly more variable than the spectral components dominating other energy bands on timescales shorter than 10\,ks. In the case of the \NXS vs \eddr we found an anti-correlation which is still not statistically significant that became less stronger in the medium and hard energy bands. 
\begin{table}
\caption{List of the best-fit relations together with their p-values for the \NXS in the soft, medium and hard energy bands. The fits are performed in the log-log space using equation Eq. \ref{eq:fitting_rel}. }
\label{tab:relations_smh}
\begin{tabular}{lcccc}
\hline 
\hline
Relation&Intercept (A) &Slope (B) &Pearson&1-P$_{\rm value}$\\
\hline
\hline
\noalign{\medskip}
\NXS$_{\rm soft}$ vs \mass & $1.23 \pm 1.08$ & $-0.68 \pm 0.16$ & -0.61 & 0.79\\
\NXS$_{\rm med}$ vs \mass  & $2.58 \pm 1.45$ & $-0.73 \pm 0.25$ & -0.87 & 0.98\\
\NXS$_{\rm hard}$ vs \mass & $2.67 \pm 1.32$ & $-0.79 \pm 0.28$ & -0.89 & 0.98\\
\noalign{\medskip}
\NXS$_{\rm soft}$ vs \lx   &$16.65 \pm 5.06$& $-0.44 \pm 0.11$ & -0.85 & 0.98\\
\NXS$_{\rm med}$ vs \lx    &$12.06 \pm 6.00$& $-0.34 \pm 0.14$ & -0.71 & 0.92\\
\NXS$_{\rm hard}$ vs \lx   &$18.06 \pm 5.64$& $-0.47 \pm 0.13$ & -0.83 & 0.99\\
\noalign{\medskip}
\NXS$_{\rm soft}$ vs \eddr & $-3.31\pm 0.25$ & $-0.67\pm 0.20$  & -0.34 & 0.57\\
\NXS$_{\rm med}$ vs \eddr  & $-2.93\pm 0.24$ & $-0.29\pm 0.19$  & -0.29 & 0.33\\
\NXS$_{\rm hard}$ vs \eddr & $-2.77\pm 0.23$ & $-0.44\pm 0.19$  & -0.15 & 0.29\\
\hline
\end{tabular}
\end{table}


\bsp	
\label{lastpage}
\end{document}